\documentclass{llncs}
\usepackage[T1]{fontenc}
\usepackage{amsmath}
\usepackage{graphicx,verbatim}
\usepackage{multirow} 
\usepackage{amssymb}

\makeatletter
\def\@oddfoot{\hfil\thepage\hfil} 
\def\@evenfoot{\hfil\thepage\hfil} 
\makeatother

\begin{document}
\title{Reconsider the Template Mesh in Deep Learning-based Mesh Reconstruction}
%

\author{Fengting Zhang\({}^{1}\), Boxu Liang\({}^{1}\), Qinghao Liu\({}^{1}\), Min Liu\({}^{1}\)\({}^{\dagger}\),  Xiang Chen\({}^{1}\)\({}^{\dagger}\),  Yaonan Wang\({}^{1}\)}  
\institute{\({}^{1}\)Hunan University}

\footnotetext[1]{$\dagger$ Corresponding authors: \texttt{liu\_min@hnu.edu.cn}, \texttt{xiangc@hnu.edu.cn}}

\maketitle              
%

%

\begin{abstract}
Mesh reconstruction is a cornerstone process across various applications, including in-silico trials, digital twins, surgical planning, and navigation. Recent advancements in deep learning have notably enhanced mesh reconstruction speeds. Yet, traditional methods predominantly rely on deforming a standardised template mesh for individual subjects, which overlooks the unique anatomical variations between them, and may compromise the fidelity of the reconstructions. In this paper, we propose an adaptive-template-based mesh reconstruction network (ATMRN), which generates adaptive templates from the given images for the subsequent deformation, moving beyond the constraints of a singular, fixed template. Our approach, validated on cortical magnetic resonance (MR) images from the OASIS dataset, sets a new benchmark in voxel-to-cortex mesh reconstruction, achieving an average symmetric surface distance of 0.267mm across four cortical structures. Our proposed method is generic and can be easily transferred to other image modalities and anatomical structures. 

\keywords{Deep Learning  \and Mesh Reconstruction \and Cortical Mesh Reconstruction \and Template-based Mesh Reconstruction.}
\end{abstract}
\section{Introduction}

Mesh reconstruction involves creating 3D surface meshes of specific shapes from a variety of images, whether they be single-view, multi-view, or volumetric 3D images. This process is pivotal in both computer vision and medical image analysis, underpinning numerous clinical applications, including surgical planning and navigation, the analysis of organ motion, and disease diagnosis~\cite{chen2021shape}.

Mesh reconstruction plays a crucial role in medical imaging for modeling anatomical structures including brain/heart/body/hand \cite{vox2cortex,chen2021shape,xia2022automatic,lin2021end,tang2021towards}. Conventional approaches employ two paradigms: 1) Isosurfacing (e.g., marching cubes \cite{marchingcubes} and FreeSurfer \cite{fischl2012freesurfer}) extracting surfaces from volumetric data, often serving as initialization; 2) Registration-based methods \cite{lim2014automatic} deforming templates via parametric/non-parametric alignment. While isosurfacing enables rapid initialization, registration ensures anatomical consistency through contour/keypoint matching. However, both require cumbersome preprocessing (segmentation/keypoint detection) and intensive computation for deformation optimization, introducing error propagation risks and scalability limitations.



Deep learning(DL)-based methods have revolutionised medical image shape reconstruction by significantly speeding up the mesh reconstruction process, allowing for quick inference post-training. These methods typically employ registration techniques~\cite{chen2021shape}, utilising either a general sphere or an anatomical atlas mesh as templates. A standard DL reconstruction workflow includes feature extraction from input images, mapping these features to mesh vertices, and employing graph convolutional networks (GCN) for template deformation based on the mapped features. For instance, Pixel2mesh~\cite{wang2018pixel2mesh} transforms a sphere into various 3D objects guided by 2D image features. Chen et al.~\cite{chen2021shape} enhanced cardiac bi-ventricle mesh reconstruction using a specific mesh in the training dataset as the template, improving both efficiency and accuracy. To reconstruct accurate cortical mesh, Bongratz et al.~\cite{vox2cortex} deformed a smooth template including four smooth cortical sub-structures, the white matter (WM) and pial surfaces of both hemispheres, to the corresponding brain MR images.

Existing DL-based mesh reconstruction methods deform static templates but struggle with complex anatomical structures due to inadequate capture of individual details. While spherical templates suffice for simple objects \cite{wang2018pixel2mesh}, reconstructing intricate geometries like cortical/cardiac meshes requires excessive deformation from generic templates. Current approaches \cite{chen2021shape,vox2cortex,lebrat2021corticalflow,ma2022cortexode} adopt closer approximations via subject-specific templates or smoothed atlases, yet remain limited by fixed topology. The critical bottleneck lies in static templates' restricted deformation capacity, necessitating adaptive initialization. Therefore, we propose that an adaptive template could significantly enhance reconstruction accuracy by better matching the diverse and specific features of each subject. Note that, the adaptive template here is a closer approximation to the target mesh prior to the deformation block, which varies according to the features of each subject, instead of a fixed template. 

In this paper, we propose an Adaptive Template-based Mesh Reconstruction Network (ATMRN), which generates an adaptive template for each subject, instead of a fixed template, for subsequent deformation in mesh reconstruction. Specifically, a U-Net-like architecture is used to extract features, and then a mesh-decoder generates the variation (compared to the mean template) for each subject from the learned features, and thereby the adaptive template can be obtained by adding the mean template with the variation of each subject. Using adaptive template meshes for subsequent deformation, our proposed method obtains state-of-the-art (SOTA) reconstruction performance on cortical mesh reconstruction.
In summary, the contributions of this paper are threefold: (1)We designed a GCN mesh-decoder block to generate adaptive templates in mesh reconstruction, which is the first work to incorporate individualised adaptive templates in DL-based mesh reconstruction networks and can be adapted into other frameworks. (2) With the adaptive template, We design a novel mesh reconstruction network, ATMRN, for mesh reconstruction and achieve SOTA cortical mesh reconstruction in the OASIS dataset~\cite{marcus2007open}. (3))A sufficient analysis of the choice of template mesh is conducted in this paper, providing a foundational framework for future exploration in medical shape reconstruction.

\begin{figure}[h]
\begin{center}
\includegraphics[width=\textwidth]{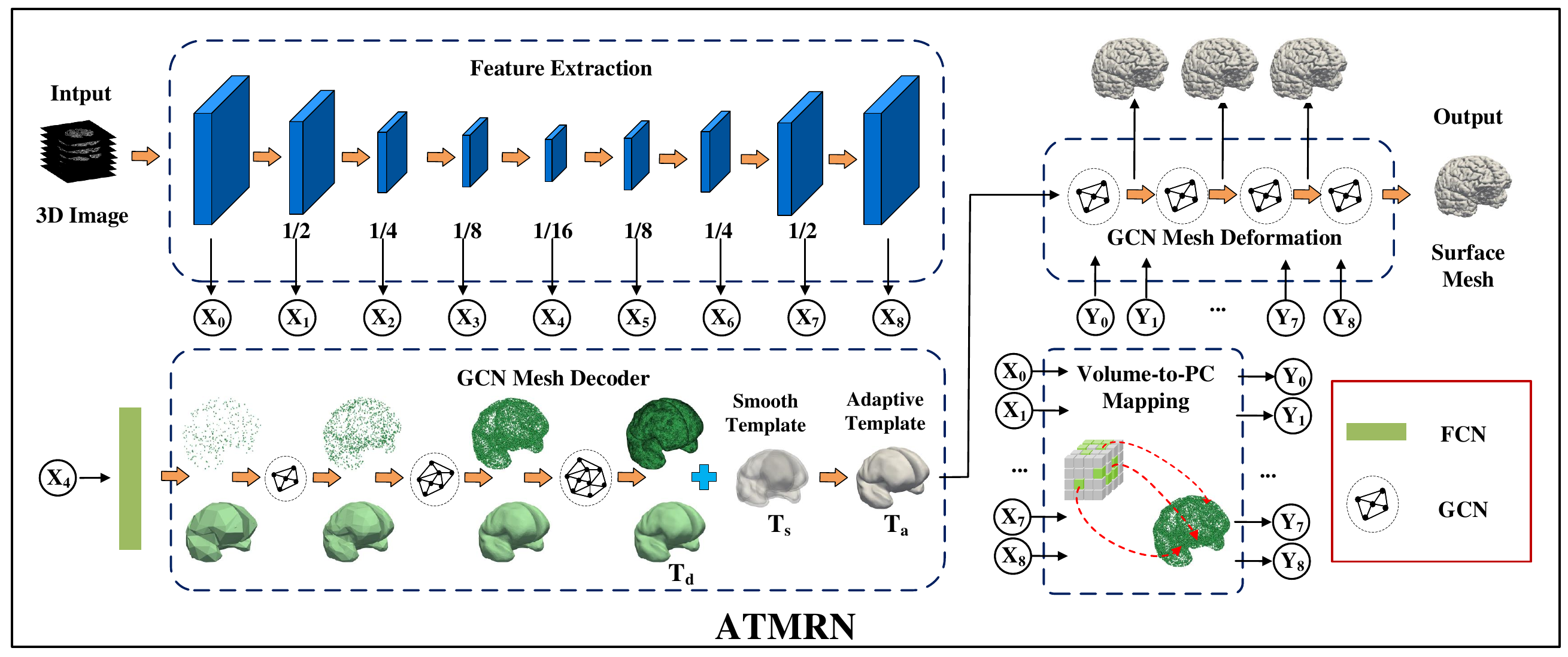}
\caption{Schema of our proposed ATMRN. Our proposed ATMRN comprises four blocks, the feature extraction, GCN mesh decoder, volume-to-PC mapping and GCN mesh deformation blocks.}
\label{fig:network}
\end{center}
\end{figure}

\section{Method}
In this section, we first introduce the network architecture of our proposed ATMRN, and then describe the corresponding loss function.

\subsection{Network Architecture}
As shown in Fig.~\ref{fig:network}, our ATMRN includes four blocks, feature extraction, GCN mesh decoder, Volume-to-point cloud (PC) mapping and GCN mesh deformation block. Given input MR images, feature extraction automatically extracts multi-resolution features and feeds them into the GCN mesh decoder and volume-to-PC mapping block. With the GCN mesh decoder, the adaptive template of each specific subject would be generated. Finally, the GCN mesh deformation block deforms the adaptive template under the guidance of mapped image features from the volume-to-PC mapping block, to achieve the target mesh.


\textbf{Feature Extraction.}
Similar to previous DL-based mesh reconstruction approaches~\cite{vox2cortex}, we utilise a U-Net-like architecture~\cite{ronneberger2015u} to extract multi-resolution features from the original images. The feature extraction block includes four downsample blocks paired with an equal count of upsample blocks, all interconnected by skip connections to ensure feature coherency across identical resolutions. Each downsample block comprises a downsample layer (convolutional layers with a stride of two), a batch normalisation layer and a ReLU activation function, halving the resolution of the input features. Correspondingly, each upsample block comprises an upsample layer(transposed convolutional layers with a stride of two), a batch normalisation layer and a ReLU activation function, doubling the resolution of the input features. Therefore, the feature extraction block learns eight multi-resolution features,$X_0$-$X_8$, respectively, where the $X_0$-$X_4$ are the encoder features and the $X_5$-$X_8$ are the decoder features.

\textbf{GCN Mesh Decoder.}
We assume the adaptive template mesh $T_{a}$ as an amalgamation of a baseline mesh $T_{s}$ and a displacement mesh $T_{d}$ (i.e. the variation of each adaptive template $T_a$ compared to the baseline mesh $T_s$, with the same shape as $T_{s}$) learned from the image features. The baseline mesh $T_{s}$ could either be a mean mesh, obtained by averaging all aligned training meshes, or a specific mesh randomly selected from the dataset (either post-smoothing or not), and the displacement mesh $T_{d}$ is learned automatically from the image feature by the GCN mesh decoder. Therefore, the generation of adaptive template mesh $T_{a}$ can be formulated as,
\begin{equation}
\label{eqn:ddf}
T_d = Dec(X_i),
\end{equation}
\begin{equation}
\label{eqn:adapative_tempalte}
T_a = T_s + T_d,
\end{equation}
where the $X_i$ is the feature maps learned from the feature extraction block and the $Dec()$ denotes the GCN mesh decoder process. Note that, in this paper, we make a simple assumption that the adaptive templates are with similar structures to the smooth template (conforming to the same topology), while having closer coordinates to the target mesh than the smooth template. For further work, the adaptive templates may also have different typologies with different faces, corresponding to more complex scenarios.

The construction of the GCN mesh decoder is inspired by the mesh autoencoder in~\cite{ranjan2018generating}. Prior to the network training, we downsampled the smooth template four times into coarse templates, leading to four new templates at different scales. In each scale, the corresponding downsampled template is a watertight smooth mesh. The corresponding downsample matrix is then used to upsample vertices from coarse to fine in the GCN mesh decoder.
Within this block, the feature $X_4$ extracted from the final encoder layer of the feature extraction module, serves as the input. Generally, more features would lead to better reconstruction accuracy. As we are the first step to explore if the adaptive templates can benefit the mesh reconstruction, a single feature from the bottom layer would be enough. A flatten operation combined with multi fully connected layers (FCN) turns the image feature into an l-dimensional latent embedding, and then FCN with reshape operation is used to generate a coarse mesh ($\frac{N}{2048} \times 3$, N is the number of vertices in target mesh) from the latent embedding. This coarse mesh undergoes a series of refinements through four GCN upsample layers, each designed to enhance mesh resolution progressively. Each GCN upsample block comprises a PC upsample layer, a graph convolution layer and an activation function, gradually turning the course mesh to successively denser meshes with $\frac{N}{512}$, $\frac{N}{128}$, $\frac{N}{16}$ and finally $N$ vertices (the displacement mesh $T_{d}$).


Note that, the generation of adaptive template conforms to the anatomical structure of cortical mesh by using fixed upsampling rules. Furthermore,  smooth regularisations are applied in the predictions of mesh deformation blocks, thereby avoiding intersections and broken meshes in the predicted meshes.
Prior studies in mesh reconstruction predominantly relied on fixed template meshes, employing either a universal sphere mesh~\cite{wang2018pixel2mesh} or structure-specific mesh~\cite{chen2021shape}. To our knowledge, this work represents the pioneering effort to adopt adaptive templates, diverging from the conventional use of a static mesh, within the domain of mesh reconstruction networks.

\textbf{Volume-to-PC Mapping.}
The features learned by the feature extraction block in the image domain cannot be directly used in the GCN deformation block, as the template mesh and image features are in different domains. The volume-to-PC mapping functions as a crucial intermediary, facilitating the transition of learned features from the image domain to the spatial domain, thereby addressing the domain discrepancy between the template mesh and the image features. We adopt the established volume-to-PC Mapping methodology in prior studies~\cite{vox2cortex}, where all image features $X_0$-$X_8$ in the image domain are mapped into spatial features $Y_0$-$Y_8$ corresponding to the adaptive template.




\textbf{GCN Mesh Deformation.}
GCN mesh deformation block deforms the adaptive template gradually into the target mesh, with the guidance of the mapped image feature $Y_0$-$Y_8$. These features are concatenated with the coordinates of the adaptive templates and fed into four GCN blocks. Different from the GCN blocks in the GCN mesh decoder, each GCN block here contains three GCN layers, batch normalisation and a ReLU activation function, where the input and output have the same vertex number. This staged approach allows for a methodical deformation of the adaptive template, enhancing detail incrementally across four stages of output meshes. Each output contributes to the computation of loss in comparison with the ground-truth meshes, facilitating a comprehensive and fine-grained adaptation to achieve a high-fidelity reconstruction.

\subsection{Loss Function.}
We employ a multi-term mesh loss function to train our proposed ATMRN. The mesh loss is designed following previous research~\cite{vox2cortex,wang2018pixel2mesh,chen2021shape}, including two parts, similarity and regularisation. The former quantifies the divergence between the geometry of the predicted mesh and that of the ground-truth, serving as a direct measure of the reconstruction's fidelity. The latter is to guarantee the topology of the reconstructed mesh is the same as the ground-truth mesh.
For similarity, we use Chamfer distance (CD) ${L}_{CD}$, which captures an overall distance between the predicted vertices and vertices of ground-truth. It does not require the point number/order to be the same on the two meshes, as the ground-truth meshes generated from FreeSurfer generally exhibit a higher density of vertices. 
For the regularisation loss, we choose edge loss ${L}_{edge}$, normal loss ${L}_{normal}$ and Laplacian loss ${L}_{Laplacian}$ (details can be found in~\cite{vox2cortex,wang2018pixel2mesh,chen2021shape}). Therefore, the mesh loss ${L}_{mesh}$ of each outputs can be formulated as, 
\begin{equation}
\label{eqn:mesh}
\begin{split}
{L}_{mesh} =  \lambda_1{L}_{CD} + \lambda_2{L}_{Laplacian} + \lambda_3{L}_{normal} + \lambda_4{L}_{edge}.
\end{split}
\end{equation}





\section{Experiments and Results}

\subsection{Experiments Settings}

\textbf{Dataset.}
We demonstrate our proposed method on brain MR images from OASIS-1 dataset \cite{marcus2007open}, comprising 416 subjects aged 18 to 96. The dataset is randomly split into 340, 42, and 43 subjects for training, validation, and testing, with a balance in diagnosis, age, and sex. Following~\cite{vox2cortex}, all the brain MR images are preprocessed (aligning to the MNI152 space, padding, cropping and resizing) using FreeSurfer into $128 \times 144 \times 128$ volumes. The corresponding segmentation and ground-truth cortical meshes are also generated using FreeSurfer, each sub-structure with about 40000 vertices. For fast convergence, the voxel intensities in the source images are normalised into $[0,1]$, and all the vertices in the target meshes are normalised into $[-1,1]$.



\textbf{Implementation Details.} Our ATMRN model is developed using PyTorch and trained on an NVIDIA RTX A6000 GPU machine. To train the network, we assign a learning rate of $1e^{-4}$ to the feature extraction block, and $5e^{-5}$ to the remaining components, with a batch size of 1. The latent vector in the GCN Mesh Decoder block is configured as $l=128$. The hyper-parameters $\lambda_1$, $\lambda_2$, $\lambda_3$ and $\lambda_4$ are set to 5, 0.1, 0.001 and 5 respectively. 


\begin{figure}[ht]
\begin{center}
\includegraphics[width=\textwidth]{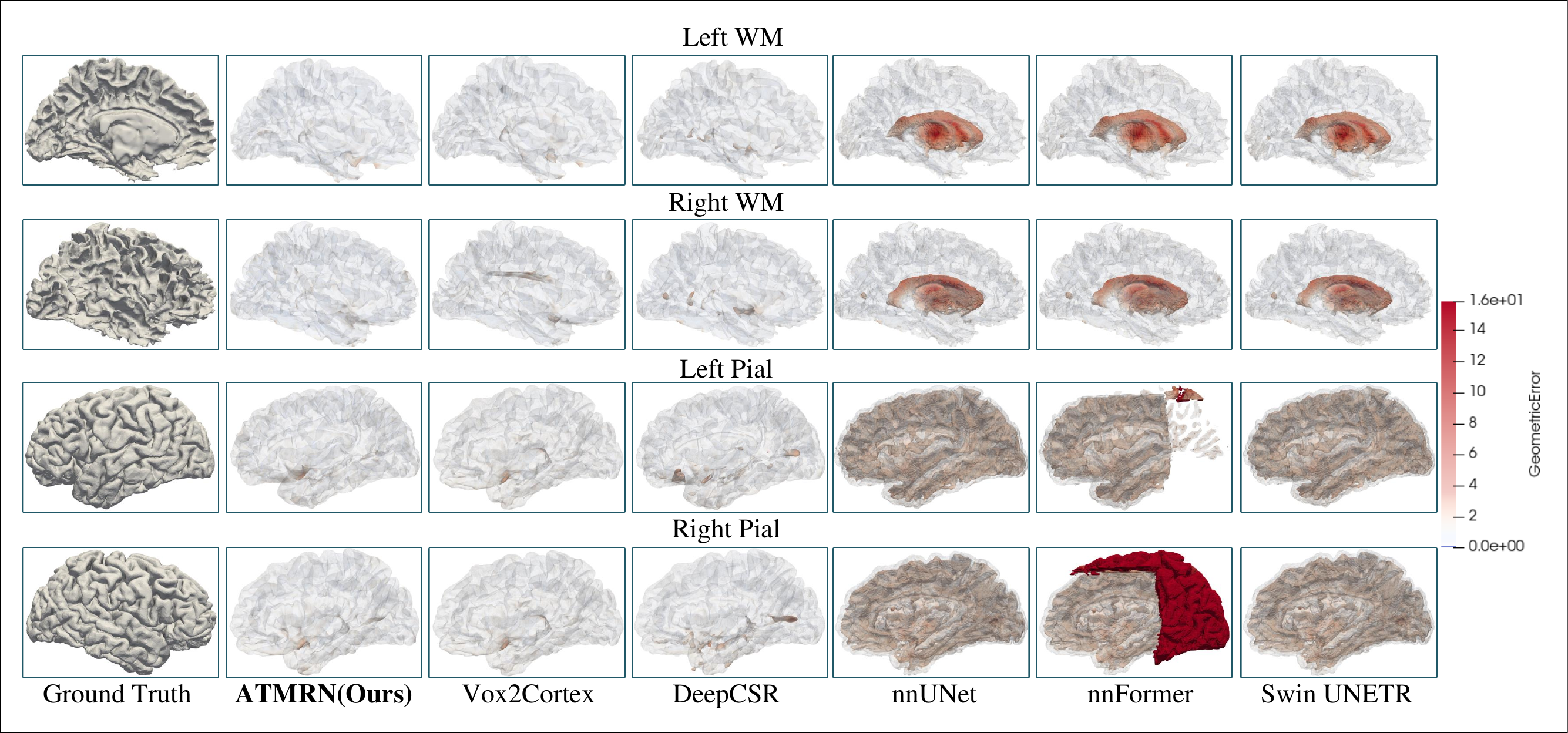}
\caption{Qualitative results between ATMRN and SOTA methods. The ground-truth meshes are on the left (from the top to bottom are the surface meshes of left WM, right WM, left pial and right pial), with the corresponding predicted meshes on the right. The colour bar denotes the distance between the predicted and ground-truth meshes.}
\label{fig:qualitative_results}
\end{center}
\end{figure}

\textbf{Baseline Methods.}
We compare our proposed method with both traditional and DL-based mesh reconstruction methods. For traditional methods, we adopt three popular DL-based segmentation methods including nnUnet~\cite{isensee2021nnu}, Swin UNETR~\cite{hatamizadeh2021swin} and nnFormer~\cite{zhou2023nnformer}, to segment the WM and pial of both hemispheres from brain MR images, and then generate 3D surface meshes using marching cubes~\cite{marchingcubes}, named as nnUnet, Swin UNETR and nnFormer, respectively.
For DL-based mesh reconstruction methods, we compare our ATMRN with two SOTA methods, the Vox2Cortex\cite{vox2cortex} and DeepCSR\cite{cruz2021deepcsr}, both trained sufficiently using the same data as our approach, for fair comparison. Following previous research~\cite{vox2cortex}, we use average symmetric surface distance (ASSD) and Hausdorff distance (HD) to evaluate the performance of mesh reconstruction.


\subsection{Results and Analysis}
\textbf{Qualitative Results.}
The qualitative comparison between our proposed ATMRN and SOTA methods is depicted in Fig.~\ref{fig:qualitative_results}, where the ground-truth meshes are on the left, with the corresponding meshes obtained by our ATMRN, Vox2Cortex, DeepCSR, nnUNet, nnFormer and Swin UNETR on the right. This side-by-side comparison clearly demonstrates the superior accuracy of DL-based methods over traditional mesh reconstruction approaches. The reconstructed surfaces by nnUNet, nnFormer and Swin UNETR are with a significant portion of vertices which are with $>2 mm$ distance from their counterparts in the ground-truth mesh (highlighted in red). Although similar discrepancies are observed in DL-based reconstruction methods, the regions are markedly reduced. Notably, the meshes predicted by our ATMRN display a consistently closer approximation to the ground-truth, with discrepancies less than 2mm across all sub-structures. 



\begin{table*}[ht]
 	\centering
 	\caption{Quantitative comparison between our ATMRN and SOTA methods, with the best performance highlighted in bold. The units for both metrics are `mm’.\label{tab:comparison}}
        \resizebox{\textwidth}{!}{
 	\begin{tabular}{c|cc|cc|cc|cc|cc}
   \hline
 		\multirow{2}{*}{Method} & \multicolumn{2}{c}{Left WM Surface} & \multicolumn{2}{c}{Right WM Surface} & \multicolumn{2}{c}{Left Pial Surface} & \multicolumn{2}{c}{Right Pial Surface} & \multicolumn{2}{c}{Average} \\
 		& ASSD $\downarrow$  & HD $\downarrow$  & ASSD $\downarrow$  & HD $\downarrow$  & ASSD $\downarrow$  & HD $\downarrow$  & ASSD $\downarrow$  & HD $\downarrow$  & ASSD $\downarrow$  & HD $\downarrow$  \\
   \hline
        nnFormer\cite{zhou2023nnformer} & 0.909 & 2.788 & 
                                         0.990 & 3.663 & 1.189 & 3.793 & 1.335 & 4.735 & 1.106 & 3.745 \\
        Swin UNETR\cite{hatamizadeh2021swin} & 0.817 & 2.568 & 
                                          0.797 & 2.567 & 1.122 & 3.798 & 1.124 & 3.745 & 0.965 & 3.169 \\
        nnUNet\cite{isensee2021nnu} & 0.705 & 1.686 & 
                                    0.675 & 1.594 & 1.031 & 3.080 & 1.027 & 3.041 & 0.859 & 2.351 \\
   \hline
 	Vox2Cortex\cite{vox2cortex} & 0.355 & 0.811 & 
      0.361 & 0.816 & 
      0.455 & 1.187 & 
      0.462 & 1.207 & 
      0.408 & 1.005 \\
    DeepCSR\cite{cruz2021deepcsr} & 0.315 & 0.666 
    & 0.341 & 0.677
    & 0.376 & 0.840
    & 0.381 & 0.872 
    & 0.353 & 0.764 \\
   \hline
ATMRN(our) & \textbf{0.218} &\textbf{0.466} 
        & \textbf{0.222} & \textbf{0.478} & \textbf{0.309} & \textbf{0.732} & \textbf{0.318} & \textbf{0.758} & \textbf{0.267} & \textbf{0.609} \\
   \hline
 	\end{tabular}}
\end{table*}

\textbf{Quantitative Results.}
The quantitative analysis reinforces the findings in the qualitative evaluation, as shown in Table~\ref{tab:comparison}. The DL-based approaches perform consistently significantly better than traditional methods. Specifically, meshes reconstructed using traditional approaches such as nnFormer, Swin UNETR, and nnUnet are with $>0.7$ mm ASSD to ground-truth. In contrast, Vox2Cortex achieves an ASSD of 0.408 mm, substantially outperforming traditional methods. DeepCSR further enhances the benchmark with an ASSD of 0.353 mm against ground-truth, showcasing its efficacy over Vox2Cortex. Our ATMRN achieves an ASSD of 0.267 mm—0.086 mm improvement over the sub-optimal method ($p<0.001$)—and an HD of 0.609 mm (0.155 mm improvement). 

\textbf{Ablation Study.} To investigate the contributions of distinct components in our method, we build two variations of ATMRN: \#v1, which omits the image decoder from the feature extraction block, and \#v2, which integrates segmentation loss
(following~\cite{vox2cortex})—computed the cross-entropy between the feature extraction block outputs and ground-truth segmentation-into the training of ATMRN. The corresponding results can be found in Table~\ref{tab:comparison}. The analysis reveals that incorporating segmentation loss adversely affected reconstruction performance, suggesting that the optimisation towards segmentation accuracy may detract from the network's ability to accurately reconstruct mesh surfaces. Conversely, the removal of the image decoder from the feature extraction block, highlights its critical role in the reconstruction process, which ensures the extraction of relevant, rich features necessary for effective downstream mesh deformation. 



\begin{table}[ht]
\begin{center}
\caption{
    Ablation study on different variations of ATMRN. The $T_{spe}$ denotes using a specific cortical mesh in the training dataset as the template. 
}
\begin{tabular}{ ccccccc }
\hline
Methods & Template & ImgDec & Segloss  &   Avg ASSD(mm) $\downarrow$ & Avg HD(mm) $\downarrow$ \\ 
\hline
\#v1 & $T_{a}$ & - & -  & 0.329 & 0.721  \\
\#v2 & $T_{a}$ & \checkmark & \checkmark  & 0.281 & 0.640 \\
\#v3 & $T_{spe}$ & \checkmark & -  & 0.355 & 0.856 \\
\#v4 & $T_{s}$ & \checkmark & -  & 0.285 & 0.673 \\
\#v5 & $T_{d}$ &\checkmark & -  & 0.526 & 1.550 \\
\#v6 & $T_{spe}+T_{d}$ &\checkmark & -  & 0.314 & 0.766 \\
our & $T_{a}$ &\checkmark & -  & 0.267 & 0.609  \\
\hline
\end{tabular}
\end{center}
\label{tab:ablation}
\end{table}

\textbf{Analysis of Template.} Template selection critically impacts DL-based mesh reconstruction performance. We examine various template variations in our ATMRN, to further demonstrate the superiority of our proposed adaptive template. As shown in Table~\ref{tab:comparison}, \#v3, \#v4, \#v5 mean using a specific cortical mesh, a smoothed cortical template (following~\cite{vox2cortex}) and the displacement mesh $T_d$ predicted by the GCN decoder, as the template for subsequent deformation. \#v6 denotes replacing the smooth template in our ATMRN with a specific template. Experiments show a smoothed template leads to better reconstruction performance than a specific template in the training dataset. Direct utilisation of the $T_d$ for subsequent deformation exhibited sub-optimal performance, highlighting constraints imposed by a low-dimensional embedding space (128 dimensions). However, significant performance enhancements are observed when the $T_d$ was combined with either a smoothed or a specific template, suggesting the efficacy of $T_d$ in capturing local variations to an existing mesh structure. 

\section{Conclusion}
We propose a novel mesh reconstruction framework, which utilises an adaptive template instead of a fixed template for subsequent deformation. Specifically, we designed ATMRN, a pioneering approach in mesh reconstruction that employs a GCN mesh decoder to predict the adaptive template for each subject. Extensive testing on the OASIS datasets validates the effectiveness of our proposed model. Through comparative analyses on different mesh templates, we underscore the superior performance and advantages of utilising an adaptive template in mesh reconstruction tasks. This advancement not only demonstrates the potential of adaptive templates in enhancing mesh reconstruction accuracy but also sets a new precedent for future research in the field. 

\bibliographystyle{splncs04} 
\bibliography{ATMRN}
\end{document}